\documentclass[aps,prl,twocolumn,superscriptaddress]{revtex4-1} %linenumbers

\usepackage{graphicx}
\usepackage{MnSymbol}
\usepackage{todonotes}
\usepackage{hyperref}
\usepackage{pdfpages}

\begin{document}

\title{High-energy magnetic excitations in Bi$_2$Sr$_2$CaCu$_2$O$_{8+\delta}$: Towards a unified description of its electronic and magnetic degrees of freedom}

\author{M. P. M. Dean}
\email{mdean@bnl.gov}
\author{A. J. A. James}
\affiliation{Department of Condensed Matter Physics and Materials Science,Brookhaven National Laboratory, Upton, New York 11973, USA}

\author{R. S. Springell}
\affiliation{Royal Commission for the Exhibition of 1851 Research Fellow,
Interface Analysis Centre, University of Bristol, Bristol BS2 8BS, UK}
\affiliation{London Centre for Nanotechnology and Department of Physics and Astronomy, University College London, London WC1E 6BT, United Kingdom}

\author{X. Liu}
\altaffiliation{Current affiliation: Beijing National Laboratory for Condensed Matter Physics, and Institute of Physics, Chinese Academy of Sciences, Beijing 100190, China}
\affiliation{Department of Condensed Matter Physics and Materials Science,Brookhaven National Laboratory, Upton, New York 11973, USA}

\author{C. Monney}
\affiliation{Swiss Light Source, Paul Scherrer Institut, CH-5232 Villigen PSI, Switzerland}
\author{K. J. Zhou}
\altaffiliation{Present address: Diamond Light Source, Harwell Science and Innovation Campus, Didcot, Oxfordshire OX11 0DE, UK}
\affiliation{Swiss Light Source, Paul Scherrer Institut, CH-5232 Villigen PSI, Switzerland}

\author{R. M. Konik}
\author{J. S. Wen}
\author{Z. J. Xu}
\author{G. D. Gu}
\affiliation{Department of Condensed Matter Physics and Materials Science,Brookhaven National Laboratory, Upton, New York 11973, USA}

\author{V. N. Strocov}
\author{T. Schmitt}
\affiliation{Swiss Light Source, Paul Scherrer Institut, CH-5232 Villigen PSI, Switzerland}

\author{J. P. Hill}
\email{hill@bnl.gov}
\affiliation{Department of Condensed Matter Physics and Materials Science,Brookhaven National Laboratory, Upton, New York 11973, USA}

% user macros
\def\mathbi#1{\textbf{\em #1}}
\date{\today}

\date{\today}

\begin{abstract}
We investigate the high-energy magnetic excitation spectrum of the high-$T_c$ cuprate superconductor Bi$_2$Sr$_2$CaCu$_2$O$_{8+\delta}$ (Bi-2212) using Cu $L_3$ edge resonant inelastic x-ray scattering (RIXS). Broad, dispersive magnetic excitations are observed, with a zone boundary energy of $\sim$300~meV and a weak dependence on doping. These excitations are strikingly similar to the bosons proposed to explain the high-energy `kink' observed in photoemission. A phenomenological calculation of the spin-response, based on a parameterization of the ARPES-derived electronic structure and YRZ-quasi-particles, provides a reasonable prediction of the energy dispersion of the observed magnetic excitations. These results indicate a possible unified framework to reconcile the magnetic and electronic properties of the cuprates and we discuss the advantages and disadvantages of such an approach.
\end{abstract}
\pacs{}
%\keywords{}
\maketitle
Of all the high-$T_c$ cuprate superconductors, Bi-2212 is the most easily cleaved to reveal atomically flat surfaces and is the preferred material for many angle-resolved photo-emission (ARPES) and scanning tunneling spectroscopy studies of the cuprates. Much is known, therefore, about the electronic structure of this cuprate. In particular, ARPES studies have revealed arcs of quasiparticle spectral weight at the Fermi energy \cite{Marshall1996,Ding1996,Loeser1996}, which on closer examination have been described as pockets \cite{Yang2008}. These have strong spectral intensity at the front surface (the `arc') and negligible spectral intensity on the back surface \cite{Yang2008,Yang2011}. This behavior was predicted by the Yang-Rice-Zhang (YRZ) ansatz as corresponding to pieces of Fermi surface characterized by the (front surface) poles, and (`ghost' surface) zeros, of the single particle Green's function. \cite{Yang2006}.

 In contrast to the electronic structure, however,  there is scant information about the  high-energy magnetic response ($>100$~meV) in Bi-2212. This is because Bi-2212 is challenging to grow as large single crystals, which are required for inelastic neutron scattering experiments. Thus only a few studies of the magnetic excitations in the cuprates have opted to measure Bi-2212 \cite{Fong1999,He2001,Fauque2007,Capogna2007,Xu2009} and these have focused around the antiferromagnetic ordering wavevector $(0.5,0.5)$, and on energy transfers below about 80~meV. The lack of information on the high energy magnetic response and its momentum dependence away from $(0.5,0.5)$ of Bi-2212 is particularly unfortunate, given the well characterized electronic structure of this cuprate. Such a measurement is also highly relevant for theoretical proposals for magnetically-mediated high-$T_c$ superconductivity. Many such theories require the existence of magnetic modes with significant spectral weight over a large fraction of the Brillouin zone in the optimally doped cuprates \cite{[{For a review see: }] Eschrig2006, *Scalapino2012}.

The dramatic improvement in recent years of the energy resolution of Cu $L_3$ edge resonant inelastic x-ray scattering (RIXS) \cite{Braicovich2009,Braicovich2010,LeTacon2011,Schlappa2012,Dean2012,Zhou2013} means that RIXS now provides a useful complement to neutron scattering. It is particularly well suited to very small sample sizes $<100$~$\mathrm{\mu}$m and can access high energy magnetic excitations above 100~meV.

This letter reports the first measurements of the high energy magnetic excitations in Bi-2212. We observe magnetic excitations with a zone boundary energy of $\sim$300~meV in both underdoped and optimally doped Bi-2212, much like a broadened remnant of the magnon excitation in undoped cuprates such as La$_2$CuO$_4$ \cite{Coldea2001,Headings2010}. These excitations are strikingly similar to the boson proposed to explain the high-energy `kink' observed in photoemission \cite{Valla2007,Graf2007}. We then take a parameterization of the measured electronic structure of Bi-2212 \cite{Yang2011} and use it to predict the magnetic excitation spectrum in an approach based on the Yang-Rice-Zhang ansatz for the electronic structure \cite{Yang2006}. Such an approach provides a reasonable prediction of the energy dispersion of the observed magnetic excitations.

\begin{figure}
\includegraphics[width=1.0\linewidth]{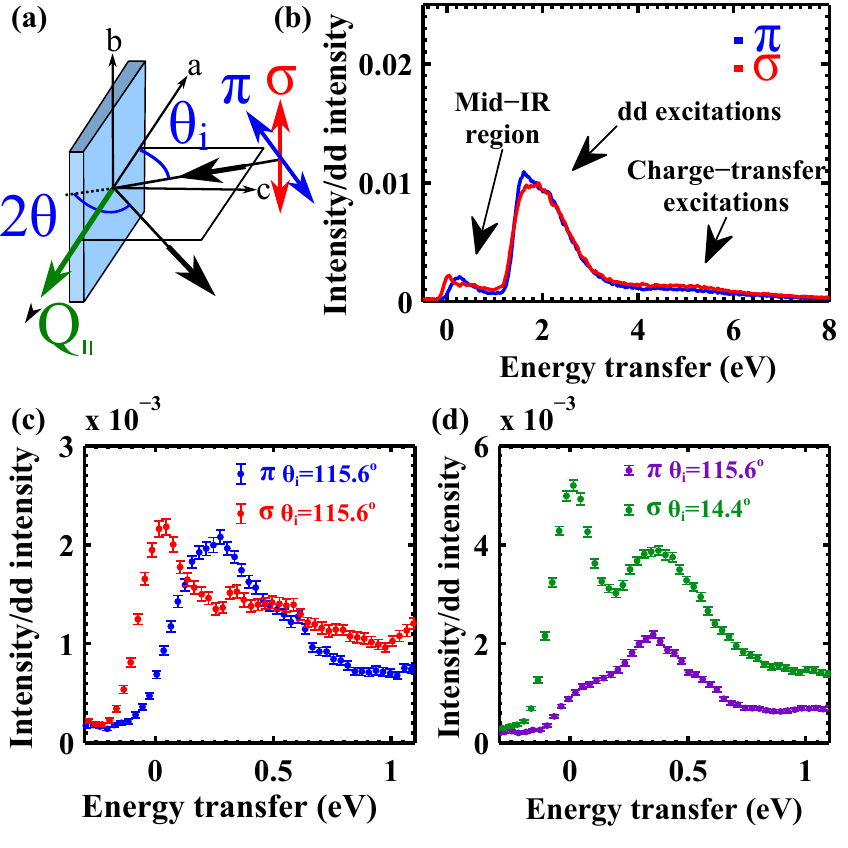} %
\caption{(Color online) (a) The experimental geometry. (b) A typical RIXS spectrum of optimally doped Bi-2212 (OD92~K). (c) OD92~K sample at $\mathbi{Q} = (0.40,0.00)$ measured with grazing-exit geometry ($\theta_i= 115.6^{\circ}$) and $\pi$ or $\sigma$ polarized light. (d) Comparison of grazing-exit ($\theta_i= 115.6^{\circ}$) $\pi$-polarized spectrum and with grazing-incidence ($\theta_i= 14.4^{\circ}$) $\sigma$-polarized spectrum for the UD0~K sample at $\mathbi{Q} = (0.40,0.00)$. We follow previous
conventions \cite{BraicovichPRB2010} and present spectra normalized to the integrated
intensity of the dd-excitations.}
\label{fig1}
\end{figure}
We prepared optimally doped Bi-2212 with $T_C = 92$~K (OD92~K) using the floating zone method \cite{Wen2008}. These crystals were heated under vacuum to produce underdoped samples $T_c<2$~K (UD0~K). Following previous work \cite{Yang2011}, we estimate the hole doping of these samples as $x=0.16$ and $x=0.03$ respectively.  Samples prepared in this way were previously studied using ARPES and showed well-defined electronic states, attesting to the high sample quality \cite{Yang2011}. As a precaution we cleaved the samples \emph{in-situ} in ultra high vacuum $\sim10^{-9}$~mbar immediately before performing the experiments. RIXS experiments were performed using the SAXES instrument at the ADRESS beamline at the Swiss Light Source \cite{Ghiringhelli2006}.  The x-ray energy was tuned to the peak in the Cu $L_3$ edge x-ray fluorescence around 931~eV. The experimental resolution of 150~meV full-width-half-maximum (FWHM), and the elastic energy, were determined by scattering from disordered graphite. We follow previous conventions \cite{BraicovichPRB2010} and present spectra normalized such that the integrated intensity of the $dd$-excitations is one. The in-plane scattering vector, \mathbi{Q}, is denoted using the pseudo-tetragonal unit cell $a=b=3.82$~\AA{} and $c=30.8 $~\AA{}, where $c$ is normal to the cleaved sample surface.  The experimental geometry is shown in Fig.~\ref{fig1}(a). X-rays are incident on the sample surface at $\theta_i$ and scattered through fixed $2\theta =130^{\circ}$. $\theta_i$ is varied in order to change $\mathbi{Q}_{||}$ -- the projection of the momentum transfer, $\mathbi{Q}$ along $[1 0 0]$.  The x-ray polarization can be chosen parallel ($\pi$) or perpendicular ($\sigma$) to the horizontal scattering plane. All data were taken at $T=15$~K.

%\section{Discussion of RIXS + Overall spectrum}
Figure~\ref{fig1}(b) plots a typical Cu $L_3$ edge RIXS spectrum of Bi-2212. We observe a relatively flat continuum of excitations up to about 8~eV, which are charge-transfer excitations from the Cu $3d$ to the O $2p$ states \cite{Ghiringhelli2004}. In the 1-3~eV range we see optically forbidden orbital $dd$-excitations where the Cu $x^2-y^2$ hole is excited into different orbitals \cite{Ghiringhelli2004}. Below 1~eV, single, and multi-spin-flip, excitations are seen. Phonons also contribute below $\sim$90~meV, along with elastic scattering at $0$~eV.

\begin{figure}
\includegraphics[width=1.0\linewidth]{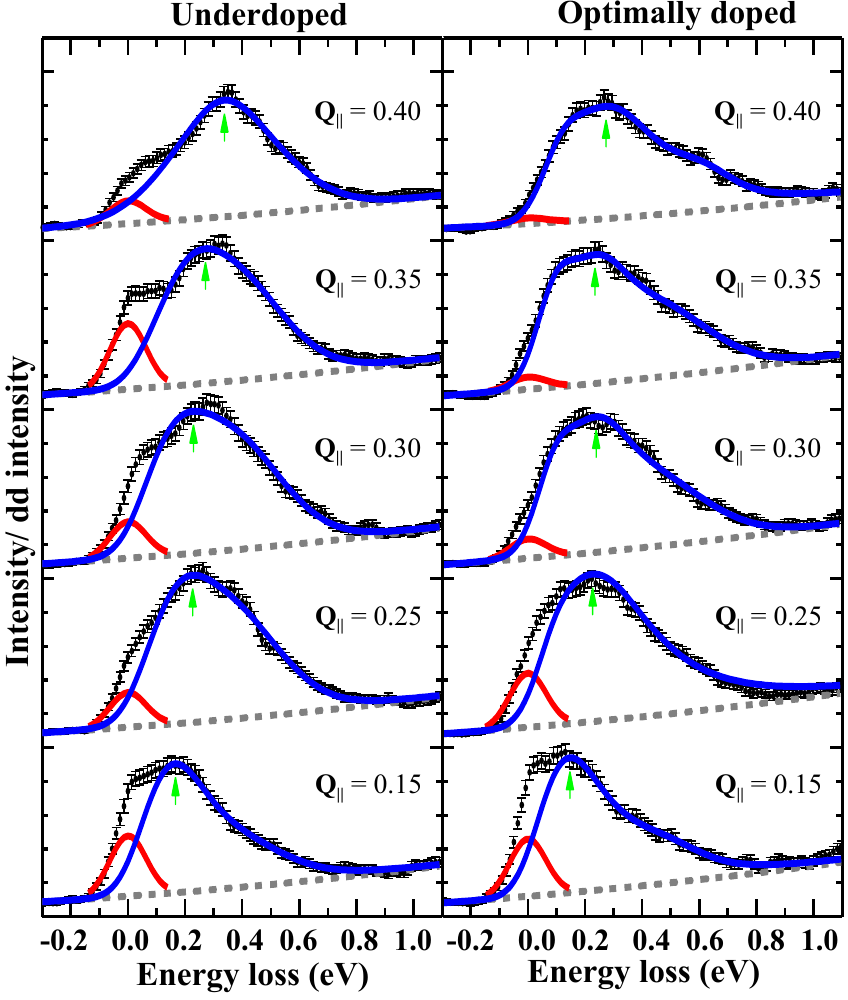}
\caption{ (Color online) The RIXS spectra of Bi-2212 along the $(0,0) \rightarrow (0.5,0)$ symmetry direction (black dots). Left: heavily underdoped non-superconducting (UD0~K); Right: optimally doped $T_C = 92$~K (OD92~K). The spectra are decomposed into the dispersive inelastic magnetic scattering (blue line), the elastic scattering (red line) and the charge-transfer/ $dd$-excitations background (gray dotted line). The data were collected at $T = 15$~K with grazing-exit geometry ($\theta_i > 65^{\circ}$) and $\pi$-polarized incident x-rays. }
\label{fig2}
\end{figure}

%\section{Polarization dependence}
The peak in the mid-IR portion of the spectrum occurs around 300~meV, near the Brillouin zone boundary. This is a comparable energy scale to the magnetic excitations seen in other doped cuprates such as La$_{2-x}$Sr$_x$CuO$_4$ \cite{Vignolle2007,Lipscombe2009,Braicovich2010}, and the YBCO and NdBCO familes of cuprates \cite{LeTacon2011, Stock2010}, suggesting that this peak corresponds to single magnon scattering. We test this assertion by examining the polarization dependence of this peak \cite{Ament2009, Haverkort2010, Igarashi2011}.  In Ref.~\onlinecite{Haverkort2010}'s formalism the dominant term in the intensity of spin-flip scattering can be written as $ I \propto | \boldsymbol{\epsilon}_o^* \cdot \mathbf{A}_{m,n} \cdot \boldsymbol{\epsilon}_i |^2$, where $\boldsymbol{\epsilon}_i$ and $\boldsymbol{\epsilon}_o$ are the ingoing and outgoing x-ray polarizations, respectively, and $\mathbf{A}_{m,n}$ is a $3 \times 3$ tensor \cite{Haverkort2010}. Assuming a Cu$^{2+}$ ion in tetragonal symmetry, $\mathbf{A}_{m,n}$ is zero except for $\mathbf{A}_{1,2}$ and $\mathbf{A}_{2,1}$.  These rules work well for describing magnon scattering in insulating cuprates including La$_2$CuO$_4$ \cite{BraicovichPRB2010}, Sr$_2$CuO$_2$Cl$_2$ \cite{Guarise2010}, and NdBaCuO$_6$ \cite{LeTacon2011}. In near-grazing-exit geometry these rules predict that the magnon should be enhanced in $\pi$ polarized x-rays and suppressed for $\sigma$ polarized x-rays. In Fig.~\ref{fig1}(c) the $\sim300$~meV peak indeed follows this polarization dependence. Further, the elastic line at 0~eV is suppressed in the $\pi$ channel, consistent with expectations for non-resonant charge scattering. For a near-grazing incidence geometry the magnon intensity is enhanced for $\sigma$ polarized x-rays and Fig.~\ref{fig2}(d)  shows a magnon peak at 300~meV, supporting the proposed contribution from spin-flip scattering.

In Fig.~\ref{fig2} we examine the $\mathbi{Q}$-dependence of the OD92~K and UD0~K spectra along the $(0,0) \rightarrow (0.5,0)$ symmetry direction using a grazing-exit configuration and $\pi$ polarized x-rays, to maximize the  spin-flip scattering and minimize the elastic line. The peak is seen to disperse to high energy near the zone boundary, which further substantiates the assignment of the peak to single spin-flip processes, as multi-spin-flip and phonon excitations would be expected to show different dispersions \cite{Hill2008, Bisogni2012}. We find that the peak is significantly broader than the experimental resolution (which is 150~meV). Thus, although magnon-like excitations persist in Bi-2212, they are significantly damped. A similar paramagnon excitation was also recently reported in other doped cuprates Nd$_{1.2}$Ba$_{1.8}$Cu$_3$O$_7$, YBa$_2$Cu$_3$O$_{6.6}$, YBa$_2$Cu$_4$O$_{8}$ and YBa$_2$Cu$_3$O$_7$ \cite{LeTacon2011}.

In order to analyze the spectra, we decompose it into its different components. The smooth background underneath the peak, which is largely independent of $\mathbi{Q}$, is from charge-transfer processes and the tail of the $dd$-excitations. At zero energy transfer a small amount of elastic scattering is visible, although this is strongly suppressed with $\pi$ polarized x-rays. Phonons contribute below $\sim 90$~meV and multi-magnon processes contribute to the high energy tail above the peak \cite{Hill2008}. The phonon and multi-spin-flip contributions can sometimes be separated from the single spin flip contribution in undoped cuprates for which the single spin-flip excitation is a well-defined, resolution-limited magnon \cite{Braicovich2010,BraicovichPRB2010,LeTacon2011,Dean2012}.  In doped cuprates, however, the peak broadening makes this difficult so, although the peak in scattering can be assigned to spin flip scattering, the contributions of phonons and multi-spin-flip processes cannot be unambiguously separated.

\begin{figure}
\includegraphics[width=1.0\linewidth]{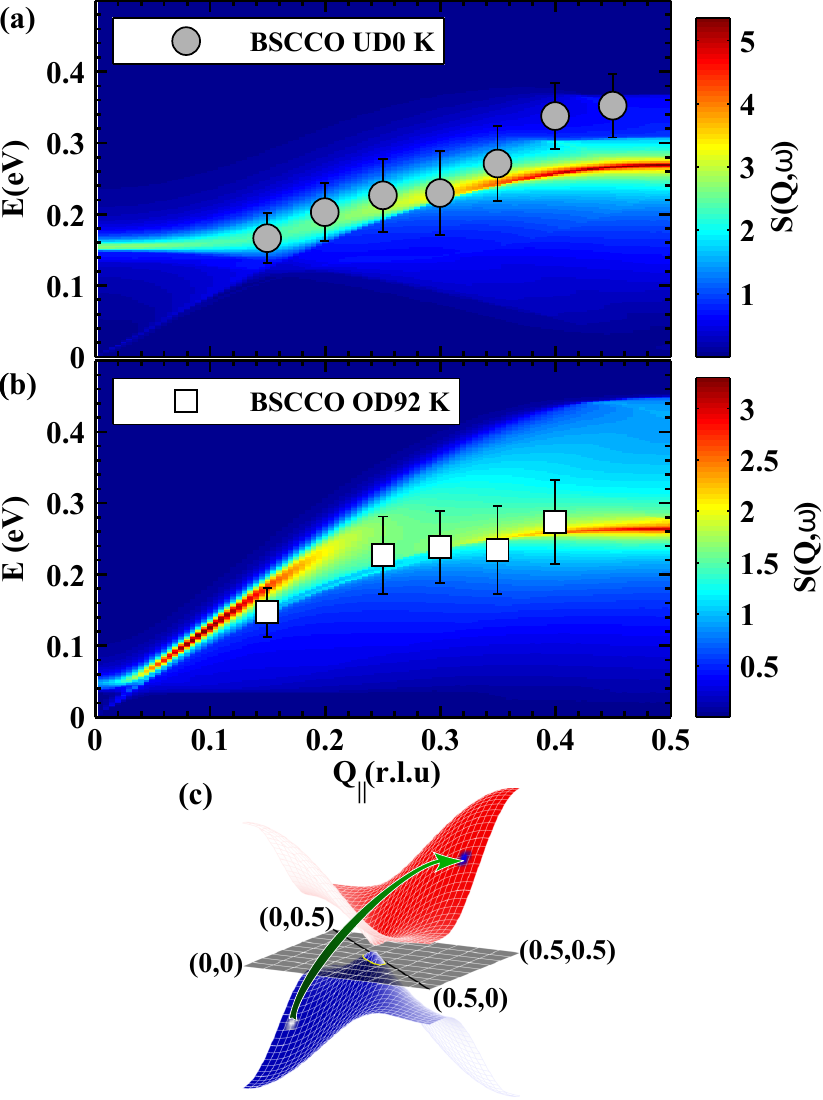}
\caption{(Color online) Comparison between the peak in the RIXS spectra dispersion (solid points) and the theoretical calculations of $S(\mathbi{Q},\omega)$ based on ARPES-derived parameters \cite{Yang2011} for (a) underdoped non-superconducting and (b) optimally doped $T_C = 92$~K Bi-2212. The error bars represent in uncertainty in determining the experimental paramagnon energy. (c) The YRZ form for the electronic structure of underdoped Bi-2212.}
\label{fig3}
\end{figure}
We track the peak in the magnetic scattering as denoted by the green arrows in Fig~\ref{fig2} (see the supplementary materials for further details) and plot the dispersion in Fig~\ref{fig3} for the (a) UD0~K and (b) OD92~K samples. The resulting dispersion shows the central result of this paper: that high energy spin excitations persist from $(0,0)\rightarrow(0.4,0)$ in Bi-2212 and are relatively unchanged on increasing the hole concentration from underdoped to optimally doped.

 There are a large number of different theoretical approaches for calculating the spin response of doped cuprates, usually starting with the Hubbard model, the $t-J$ model, or models based on stripe correlations. These models are then solved by mean field techniques or calculations based on small clusters. %
%\todo[inline]{Does this need citations? \cite{Georges1996} \cite{Vojta2009} \cite{LeTacon2011} }
%Such a phenomenology is, in fact, reproduced in calculations of $S(\mathbi{Q},\omega)$ from the Hubbard model \cite{scalapino2007}.
One difficulty is that few of these approaches allow for calculations of the electronic \emph{and} magnetic properties using a single set of parameters. For example, it has been argued that local moments contribute to the spin response at low energies \cite{Xu2009} and local moment stripe-based theories provide a very natural description of the `hourglass' feature seen in the magnetic response near $(0.5,0.5)$ \cite{Vojta2009}, but it is difficult to connect these theories with the itinerant behavior seen by ARPES.

The YRZ phenomenology \cite{Yang2006}, which was inspired by calculations on arrays of coupled two leg ladders \cite{Konik2006}, is advantageous in that it provides a relatively simple explanation for ARPES data and an easy way to implement parametrization for the electronic structure of the cuprates. It was used in just this way for underdoped Bi-2212 \cite{Yang2011}. Figure~\ref{fig3}(c) plots this electronic structure for UD0~K Bi-2212 ($x=0.03$).
It has also proved successful in describing many other \emph{electronic} properties of the cuprates \cite{Rice2012}.

In Ref.~\cite{James2012} it was argued that calculations based on the quasiparticle band structure given by YRZ are also capable of describing the magnetic response of the cuprates. Here we test this assertion by comparing the magnetic response from YRZ, using parameters set by ARPES experiments, with RIXS data on samples made in the same way. The approach taken in Ref.~\cite{James2012} was to set YRZ in the framework of earlier calculations by Brinckmann and Lee \cite{Brinckmann2001} for $S(\mathbi{Q},\omega)$, based on a slave boson mean field treatment of the $t-J$ model. The key feature suggested by the success of YRZ is that one must go beyond mean field theory and take into account the binding between fermionic `spinons' and bosonic `holons'. The basic element is a particle-hole `bubble' diagram of YRZ propagators. A geometric series of such diagrams is then summed to infinite order leading to an `RPA' expression for the magnetic response. (See the supplementary materials for a technical description.)
The parameters in this phenomenological approach were fixed by fits of the YRZ model to ARPES measurements on Bi-2212 crystals \cite{Yang2011}. The fits assume a pseudogap that decreases linearly with doping and vanishes in the overdoped regime, consistent with other ARPES studies \cite{Yoshida2012}. These parameters were then used to calculate $S(\mathbi{Q},\omega)$. We find a broad paramagnon at high energies. We also note that a highly intense mode arises around $\mathbi{Q} = (0.5,0.5)$ with an energy of $\sim~40$~meV \cite{James2012}, similar to the `resonance mode' seen in neutron scattering \cite{Fong1999,He2001,Fauque2007,Xu2009}. Unfortunately, this is above the maximum $|\mathbi{Q}|$ accessible in Cu $L_3$ edge RIXS.

Figure \ref{fig3} compares the results of these calculations to the peak in the magnetic response observed for (a) UD0~K and (b) OD92~K samples. The calculations reproduce the energy dispersion, except for a small discrepancy in the UD0~K sample at high $\mathbi{Q}$. The overall level of agreement in the energy of the paramagnon excitation is encouraging given that the calculations are entirely constrained by the electronic structure measurements \cite{Yang2011}.

We note that a comparison of the experimental and theoretical widths is difficult to perform in a meaningful way, as the width of the measured spectrum is dominated by the experimental resolution (150~meV) and the width is increased by the contribution from phonons and multi spin flip processes. Also, YRZ makes an ansatz for the coherent part of the single particle Green's function and neglects the incoherent piece. The latter takes the form of a large `hump' in ARPES data suggesting that in fact lifetime effects play an important role. A more sophisticated theory, in which this effect is accounted for, would probably predict larger widths.

Experimentally, we find similar intensities in the UD0~K and OD92~K samples, although it is difficult to measure RIXS on an absolute intensity scale. Neutron scattering measurements observe a strong suppression of spectral weight around $(0.5,0.5)$ in doped cuprates \cite{Fujita2012NS}. Taken together the x-ray and neutron results then suggest a strong $\mathbi{Q}$-dependence to the doping dependence, with spectral weight largely conserved, except around $(0.5,0.5)$. Such a $\mathbi{Q}$-dependence can be reproduced within YRZ \cite{James2012}, and other theoretical approaches such as Ref.~\cite{scalapino2007}.  Upon increasing the doping from heavily underdoped to optimally doped, the calculations predict a small drop of 15\% in the intensity of the paramagnon. In future theoretical work we intend to include the possible appearance of an ordered moment in the YRZ description at low doping. This should improve the accuracy of the theory for $\mathbi{Q}\rightarrow (0,0)$, which is currently dominated by the large pseudogap.

Our results also have relevance to ARPES studies. Several measurements of Bi-2212, and other cuprates, observe a kink of the electronic bands in the mid-IR energy scale \cite{Valla2007,Graf2007,Inosov2007}. In Bi-2212 this feature occurs around 340~meV and it is weakly doping dependent \cite{Valla2007}. These characteristics are similar to the magnetic excitations near the Brillouin zone boundary reported here, and indeed, calculations \cite{scalapino2007} have shown that a paramagnon with just the characteristics of the mode we measure would produce the observed phenomenology of the high energy kink. However, it remains unclear whether the high energy kink is due to electron-boson coupling, or to other effects \cite{Inosov2007,Byczuk2007}.

In conclusion, we have used RIXS to determine the high energy magnetic excitation spectrum of Bi-2212. We find broad dispersive paramagnons with a zone boundary energy of $\sim$300~meV and a weak dependence on doping. The observed energy dispersion of these excitations are reasonably well reproduced by calculations based on the YRZ model \cite{Yang2006} with parameters fixed by the measured electronic structure of Bi-2212 \cite{Yang2011}. This indicates that this is a promising path towards a possible unified description of the magnetic and electronic properties of the cuprates, as opposed to models that focus on just one of these degrees of freedom and we discuss the advantages and disadvantages of such an approach. Future work should improve the limitations of the current theory and test whether other spectroscopies can be incorporated into the same unified picture.

\section{Acknowledgments}
We thank H.-B. Yang, J. D. Rameau, P. D. Johnson and T. M. Rice for valuable discussions and P. D. Johnson and W.-G. Yin for preliminary calculations. M.P.M.D., A.J.A.J., R.M.K., Z.X., G.D.G.\ and J.P.H.\ are supported by Center for Emergent Superconductivity, an Energy Frontier Research Center funded by the U.S.\ DOE, Office of Basic Energy Sciences. Work at Brookhaven National Laboratory was supported by the Office of Basic Energy Sciences, Division of Materials Science and Engineering, U.S. Department of Energy under Award No.\ DEAC02-98CH10886. C.M., K.J.Z, and T.S.\ acknowledge support from the Swiss National Science Foundation and its NCCR MaNEP. The experiment was performed at the ADRESS beamline of the Swiss Light Source using the SAXES instrument jointly built by Paul Scherrer Institut, Switzerland and Politecnico di Milano, Italy. Preliminary measurements were performed at the X1A2 and X22C beamlines at the National Synchrotron Light Source, Brookhaven National Laboratory, which is supported by the US DOE under Contract No. DE-AC02-98CH10886. We thank Stuart Wilkins for support at the X1A2 beamline.

\newpage
\thispagestyle{plain}
\mbox{}
\newpage
\thispagestyle{plain}
\mbox{}
\includepdf[pages=1]{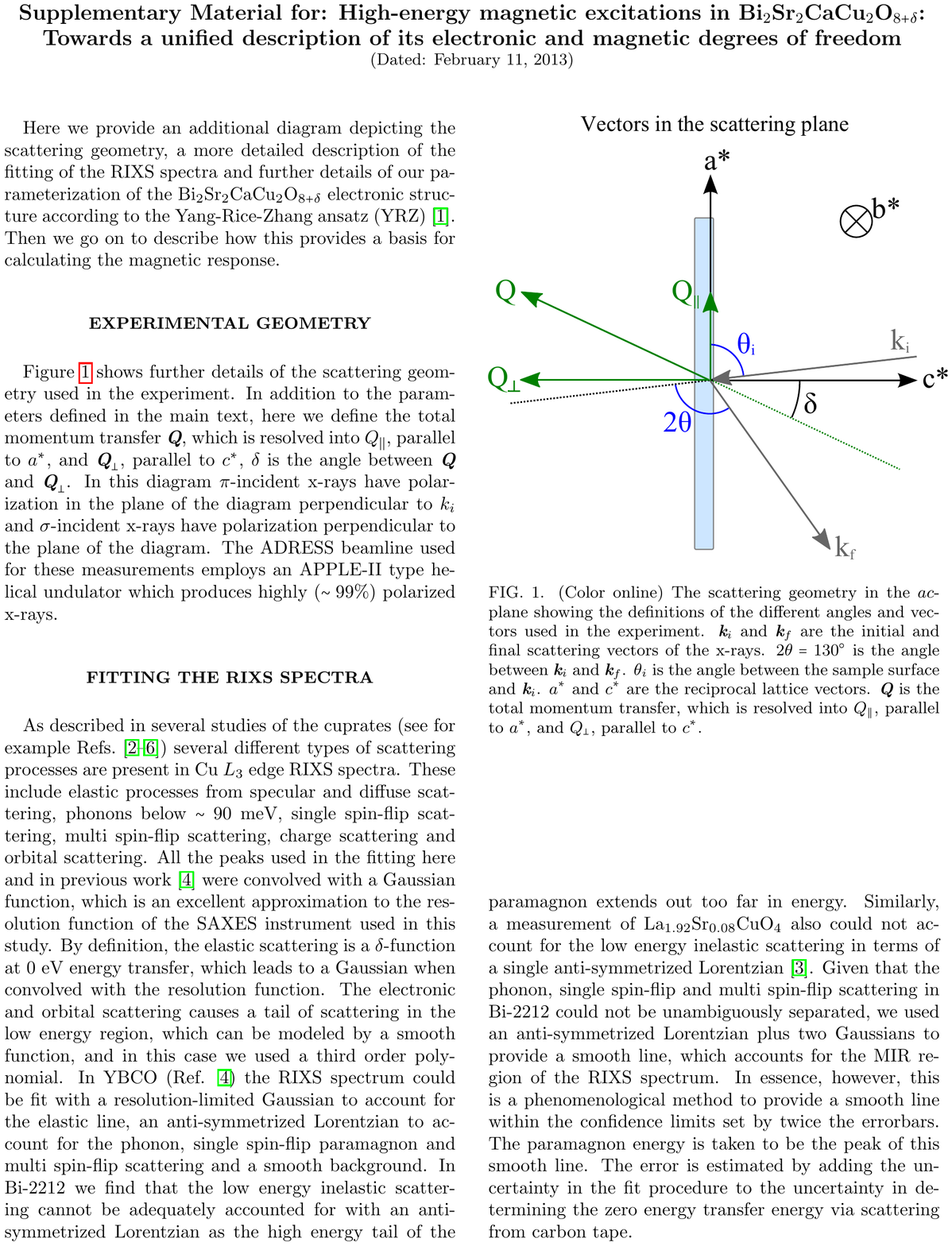}
\newpage
\thispagestyle{plain}
\mbox{}
\includepdf[pages=2]{supplementary.pdf}
\newpage
\thispagestyle{plain}
\mbox{}
\includepdf[pages=3]{supplementary.pdf}
\newpage
\thispagestyle{plain}
\mbox{}
\includepdf[pages=4]{supplementary.pdf}

\end{document}